\documentclass[final]{aipproc}

\usepackage{latexsym}
\layoutstyle{6x9}

\begin{document}

\title{Low Energy Tests of the Standard Model with Spin Degrees of 
Freedom\footnote{Plenary talk presented at the 17th International Spin Physics 
Symposium (SPIN 2006), Kyoto University, Kyoto, Japan, Oct.\ 2--7, 2006.}}


\author{Jens Erler}{
address={Departamento de F\'isica Te\'orica, Instituto de F\'isica, Universidad
         Nacional Aut\'onoma de M\'exico, M\'exico D.F. 04510, M\'exico}
}

\begin{abstract}
After briefly reviewing the status of the standard model, I will focus mainly
on polarized electron scattering and other tests of the weak neutral current.
I will also address other low energy tests in which polarization degrees of 
freedom play a crucial role, including precision muon physics and searches for 
electric dipole moments.
\end{abstract}

\maketitle


\subsection{Introduction}
The Standard Model (SM) of the strong and electroweak (EW) interactions, based 
on the gauge group, $SU(3)_C\times SU(2)_L\times U(1)_Y$, is well tested up to
energies of ${\cal O}(100\mbox{ GeV})$. It is now clear that the SM is correct 
not only to first (tree-level) order, but also at the level of radiative (loop)
corrections. Thus, beyond the SM physics can only be a small perturbation, and
is probably of decoupling type. The prospects to eventually find new physics 
are extraordinarily bright. Most theorists argue that within the SM the EW 
scale is unstable, because radiative corrections would generally drive 
the quadratic term of the Higgs potential to very high mass scales 
(the hierarchy problem) --- unless those corrections are controlled by 
a physical cut-off which is not much larger than the EW scale itself.
In addition, observations of dark matter, dark energy, and the matter 
anti-matter asymmetry in the universe imply modifications of the SM beyond 
the introduction of neutrino mass.

Most scenarios for physics beyond the SM are guided by the hierarchy problem.
{\sl Supersymmetry} (SUSY) stabilizes the Higgs potential by virtue of 
non-renormalization theorems. {\sl Dynamical symmetry breaking} ({\em e.g.\/}, 
technicolor) nullifies the problem by avoiding fundamental scalar fields 
to start with. {\sl Large extra dimensions} relate the hierarchy to 
the geometry of a higher dimensional space-time, but the stability of 
the latter remains in general an open question. {\sl Little Higgs} models 
construct the Higgs as a pseudo-Goldstone boson, postponing the occurrence of 
quadratic divergences by one or two loop orders. In all cases is it difficult 
to construct realistic models which are free of problems and consistent with 
all observations. One usually needs to introduce extra degrees of freedom 
to address those difficulties. This assures a rich phenomenology with 
implications for low energy physics, as well.

A great deal of experimental information (in several cases with better than per
mille precision) has been gained from the $Z$-factories, LEP~1 and 
SLC~\cite{LEPSLD:2005em}. As a result, the $Z$ boson is now one of the best 
studied particles of the SM, and its properties are in reasonable agreement 
with the SM. Nevertheless, there are classes of new physics which do not 
significantly affect $Z$ boson properties, and which may hide under the $Z$ 
resonance. As will be reviewed in the subsequent sections, experiments at very 
low energies --- even if their relative precisions are not at the per mille 
level --- can have complementary sensitivities to new physics. The key idea is 
to exploit the spin degree of freedom to separate the dominant parity 
conserving electromagnetic force from the parity violating EW interaction, and 
possibly parity violating new interactions. If in addition, the SM prediction 
is parametrically suppressed (as frequently turns out to be the case) one has 
enhanced leverage, allowing to test new physics scales up to the multi-TeV 
region. Thus, there are generally two complementary strategies to test the SM 
and its extensions, namely using high energy or high precision. In turn, 
precision tests can be performed in SM allowed processes ({\em e.g.\/}, parity 
violating scattering) or in SM forbidden (highly suppressed) observables (such 
as permanent electric dipole moments). 

\subsection{Status of the Standard Model}
\begin{figure}[t]
  \includegraphics[height=0.3\textheight]{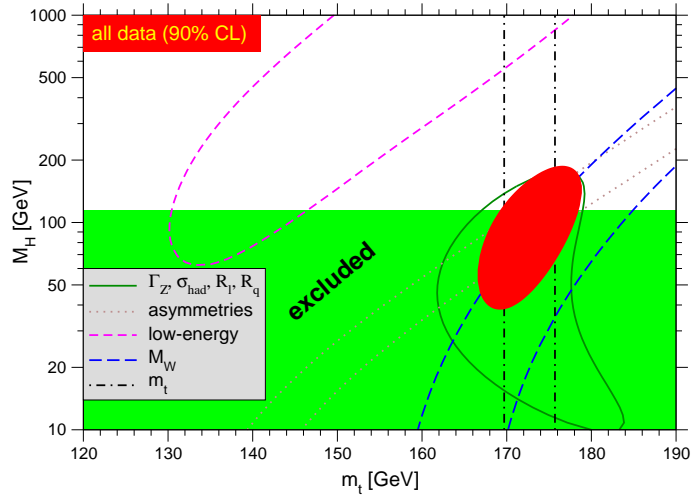}
  \caption{One-standard-deviation (39.35\%) uncertainties in $M_H$ as 
           a function of $m_t$ for various inputs, and the 90\% C.L.\ region 
           allowed by all data. The 95\% C.L.\ direct lower limit from LEP~2 is
           also shown.}
\label{fig:mhmt}
\end{figure}

One of the key parameters of the electroweak SM is the weak mixing angle,
\begin{equation}
\sin^2\theta_W = {g{'2}\over g^2 + g{'2}} = 1- {M_W^2\over M_Z^2},\hspace{50pt}
e = g \sin\theta_W = g' \cos\theta_W,
\end{equation}
where $g$, $g'$, and $e$ are the gauge couplings of $SU(2)_L$, $U(1)_Y$, and 
QED, respectively. The weak $Z^0$ boson and the photon, $A$, are then 
the linear combinations,
\begin{equation}
Z^0_\mu = \cos\theta_W W^3_\mu - \sin\theta_W B_\mu, \hspace{50pt}
A_\mu = \sin\theta_W W^3_\mu + \cos\theta_W B_\mu,
\end{equation}
of $SU(2)_L$ and $U(1)_Y$ gauge bosons, $W^3$ and $B$. Measurements of 
$\sin\theta_W$ currently yield the strongest constraints on the Higgs boson 
mass, $M_H$, which is extracted from Higgs loop effects. Fig.~\ref{fig:mhmt} 
shows that there are actually three independent determinations of $M_H$ as 
functions of $m_t$. The banana shaped solid (dark green) contour arises from 
$Z$ boson properties, like total and partial widths and the hadronic peak cross
section, but not asymmetry measurements. The latter result in the dotted 
(brown) lines. The long-dashed (blue) lines are due to the $W$ boson mass 
measurements, $M_W = 80.394 \pm 0.029$~GeV at LEP~2~\cite{LEP:2005di,ICHEP2006}
($e^+e^-$) and the Tevatron~\cite{Affolder:2000bp,Abazov:2002bu} ($p\bar{p}$). 
These three contours overlap for values of $m_t$ consistent with the Tevatron 
average (shown as the vertical lines),  
$m_t = 171.4 \pm 2.1$~GeV~\cite{Brubaker:2006xn}. Only the dashed (magenta) 
contour from low energies driven by the NuTeV result~\cite{Zeller:2001hh} on 
neutrino deep inelastic scattering ($\nu$-DIS) (to be discussed later) 
disagrees. With the latest experimental results, a global fit to all data 
yields,
\begin{equation}
\begin{array}{rcl}
M_H &=& 84^{+32}_{-25}~{\rm GeV},\\
m_t &=& 171.4 \pm 2.1~{\rm GeV}, \\
\alpha_s(M_Z) &=& 0.1216 \pm 0.0017.
\end{array}
\label{fit}
\end{equation}
The result for $M_H$ is only barely consistent (within 1~$\sigma$) 
with the 95\% C.L.\ lower search limit from LEP~2~\cite{Barate:2003sz}, 
$M_H > 114.4$~GeV. Including the results of these direct searches as an extra 
contribution to the likelihood function yields the 95\% C.L.\ upper bound,
\begin{equation}
   M_H \leq 178\mbox { GeV}.
\end{equation}
The value of $m_t$ in Eq.~(\ref{fit}) is completely dominated by the Tevatron 
input. One can also perform a fit to the precision data alone, {\em i.e.\/}, 
excluding the direct $m_t$ from the Tevatron, yielding 
$m_t = 171.0^{+9.5}_{-7.1}\mbox{ GeV}$, in perfect agreement with the direct 
determination. The strong coupling constant, $\alpha_s(M_Z)$, is mainly 
constrained by $Z$ and $\tau$ decays. These correspond to clean determinations 
of $\alpha_s(M_Z)$ at two very different energy scales and are in perfect 
agreement with each other. The overall goodness of the global fit is very 
reasonable, with $\chi^2=47.3$ for 42 degrees of freedom and a probability for 
a larger $\chi^2$ of 27\%. However, there are a few observables showing 
interesting deviations from the SM. Some of these are in the low energy sector 
and will be discussed below. 

\subsection{Effective Lepton-Hadron Lagrangian}
The weak neutral current can be tested at low energies by isolating 
interference effects with the photon amplitude using the parity (P) and charge
conjugation (C) symmetry violating nature of the weak interaction. It is 
sufficient to work with the effective P and C violating four-Fermi 
lepton-hadron Lagrangian,
\begin{equation}
{\cal L}_{\rm NC}^{lh} = {G_F\over\sqrt{2}} \sum\limits_q \left[
C_{1q}\bar{\ell}\gamma^\mu\gamma_5 \ell \bar{q}\gamma_\mu          q +
C_{2q}\bar{\ell}\gamma^\mu         \ell \bar{q}\gamma_\mu \gamma_5 q +
C_{3q}\bar{\ell}\gamma^\mu\gamma_5 \ell \bar{q}\gamma_\mu \gamma_5 q \right].
\end{equation}
Here $G_F$ is the Fermi constant, and the $C_{ij}$ are effective four-Fermi 
couplings, where,
\begin{equation}
C_{1q} = - T_3^q + 2 Q_q \sin^2\theta_W,\hspace{15pt}
C_{2u} = - C_{2d} = - {1\over 2} + 2 \sin^2\theta_W,\hspace{15pt}
C_{3u} = - C_{3d} =   {1\over 2},
\end{equation}
holds at the SM tree level. Notice, that the $C_{2q}$, as well as 
the combination $Q_W^p \equiv 2 C_{1u} + C_{1d}$ (relevant for the Qweak 
experiment discussed later), are proportional to $1 - 4\sin^2\theta_W$. 
Therefore, one can have enhanced sensitivity to $\sin^2\theta_W$ as its 
numerical value is close to 1/4. If the SM tree level contribution is 
suppressed in this way then loop effects --- but also possible new physics 
contributions --- are relatively enhanced. This gives additional leverage 
to study the TeV scale. In the following sections, past and future low energy 
measurements constraining the $C_{ij}$ are reviewed.

\subsection{PV-DIS}
The right-left asymmetry, $A_{RL}$, in parity violating deep inelastic electron
scattering (PV-DIS) is given by,
\begin{equation}
A_{RL} = {3 G_F Q^2\over 10\sqrt{2}\pi\alpha(Q^2)}\left[ (2 C_{1u} - C_{1d}) +
g(y) (2 C_{2u} - C_{2d}) \right],
\end{equation}
where $\alpha(Q^2)$ is the electromagnetic coupling at squared momentum 
transfer, $Q^2$, and $g(y)$ is a function of the fractional energy transfer, 
$y$, from the electron to the hadrons. The relative weights of up and down 
quarks is given by their electric charge ratio --- a consequence of interfering
individual quarks with the photon amplitude as is typical in the deep-inelastic
regime. The first experiment of this type was the celebrated E--122 experiment
at SLAC~\cite{Prescott:1978tm} which was crucial to establish the SM even 
before the discovery of the $W$ and $Z$ bosons (searches for atomic parity 
violation at the time gave conflicting results). The NA--004 experiment at 
CERN~\cite{Argento:1982tq} is the only experiment to date to have replaced 
positive muons with negative ones simultaneously with the reversal of the muon 
polarization. This resulted in unique sensitivity to the coefficients $C_{3j}$.

An experiment at JLab~\cite{Zheng:2006} is approved to use the current 6~GeV 
CEBAF beam to repeat the SLAC experiment on deuterium with greater precision. 
One hopes to be able to collect more data points after the 12~GeV 
upgrade~\cite{Reimer:2006}. This would improve the SLAC result and the current 
world average by factors of 54 and 17, respectively. The issues to be addressed
are higher twist effects and charge symmetry violating (CSV) parton 
distribution functions (PDFs). Since higher twist effects are strongly $Q^2$ 
dependent and CSV should vary with the kinematic variable, $x$, while 
contributions from beyond the SM would be kinematics independent, one can 
separate all these possible effects by measuring a large array of data points. 
Thus, a great deal can be learned about the strong and weak interactions at 
the same time. The measurements are expected to be limited experimentally by 
the determinations of the polarization and the $Q^2$ scale.

\subsection{Polarized M\o ller Scattering}
An experiment free of QCD issues has been completed recently by the E--158 
Collaboration~\cite{Anthony:2005pm} located in End Station A at SLAC. They 
obtained the first measurement of the parity violating M\o ller asymmetry, 
\begin{equation}
  A_{PV} = - {\cal A}(Q^2,y) Q_W^e = ( - 1.31 \pm 0.14 \pm 0.10)\times 10^{-7},
\label{moller}
\end{equation}
where ${\cal A}$ is the analyzing power and $Q_W^e$ is the so-called weak 
charge of the electron which contains all the weak physics. The experiment used
the SLC beam delivering $89\pm 4\%$ polarized electrons. The beam energies were
at 45 and 48~GeV, but the small electron mass turned this into a low 
$Q^2 = 0.026$~GeV$^2$. Although $Q_W^e$ is not in the sector giving rise to 
$C_{ij}$ measurements, it is another example of a quantity proportional to 
$1 - 4\sin^2\theta_W$. The resulting $Q_W^e = - 0.0403 \pm 0.0053$, which is in
reasonable agreement with the SM prediction, can thus be used to extract 
a precise value for the weak mixing angle defined at the $Z$ scale. Including
one-loop radiative corrections~\cite{Czarnecki:1995fw} one arrives at,
\begin{equation}
  \sin^2\theta_W (M_Z) = 0.2330 \pm 0.0014. 
\label{kk}
\end{equation}
The world's best measurements of the weak mixing angle~\cite{LEPSLD:2005em} 
have been provided by SLD ($\pm 0.00029$ from the left-right cross section 
asymmetry) and the LEP groups ($\pm 0.00028$ from the forward-backward 
asymmetry for $b$-quark final states). These two measurements contribute 
greatly to our current knowledge of $M_H$. However, they disagree from each
other by 3.1~$\sigma$. It is important to resolve this discrepancy. Notice that
their uncertainty is about 5 times smaller than the one in Eq.~(\ref{kk}). 
Thus, a factor of 5 improvement in the precision of the M\o ller asymmetry 
would make this kind of measurement fully competitive with the $Z$ factories 
which could shed some light on the discrepancy. Precisely this kind of 
improvement is currently under discussion at JLab~\cite{Mack:2006}.

\subsection{Qweak}
A very similar experiment, in fact using the same kind of target (hydrogen),
will measure the analogous weak charge of the proton, 
$Q_W^p = 2 C_{1u} + C_{1d}$. The combination of a smaller beam energy of 
1.165~GeV and the larger target mass (protons) relative to E--158 results in 
virtually the same $Q^2 = 0.03$~GeV$^2$, corresponding to elastic scattering. 
Thus, one scatters from the proton as a whole, so that the relative weight of 
up and down quarks is given by the valence quark composition. With an expected 
polarization of $85\pm 1\%$ the Qweak Collaboration anticipates to measure 
the parity violating asymmetry,
\begin{equation}
  A_{PV} = 9\times 10^{-5}\mbox{ GeV }(Q^2 Q_W^p + Q^4 B) \sim
           ( - 2.68 \pm 0.05 \pm 0.04) \times 10^{-7},
\label{apv}
\end{equation}
where the first uncertainty is experimental. The second uncertainty is from
the leading form factor contribution, the $Q^4 B$ term, which will be 
determined experimentally by means of a fit to existing and future measurements
at various $Q^2$ points. The actual Qweak experiment~\cite{Michaels:2006} is 
the one with the lowest lying $Q^2$. The anticipated errors in $Q_W^p$ and 
the corresponding $\sin^2\theta_W$ are $\pm 0.003$ and $\pm 0.0007$, 
respectively. Notice, that the (expected) asymmetry~(\ref{apv}) is about twice 
as large as the M\o ller asymmetry~(\ref{moller}). The reason is that neither 
the form factor term (suppressed by $Q^2/m_p^2$) nor the one-loop $WW$-box (of 
order $\alpha/\pi$ and enhanced by a factor of 7 relative to $Q_W^e$) enter 
with the $1 - 4\sin^2\theta_W$ suppression factor, so that on balance
these contributions are roughly comparable with the tree level. The one-loop 
radiative corrections~\cite{Marciano:1982mm} have the form,
\begin{equation}
  Q_W^p = [ \rho_{\rm NC} + \Delta_e] [1 - 4\sin^2\hat\theta_W(0) + \Delta_e']
        + \Box_{WW} + \Box_{ZZ} + \Box_{\gamma Z},
\end{equation}
a structure which also applies to $Q_W^e$, as well as to APV discussed below. 
$\rho_{\rm NC}-1$, $\Delta_e$, and $\Delta_e'$ are due to self-energy and 
vertex corrections, and $\sin^2\hat\theta_W(0)$ is an effective low energy weak
mixing angle. The $\gamma Z$-box, $\Box_{\gamma Z}$, is plagued by 
long-distance QCD effects of the form ${\ln M_Z^2/\Lambda_{\rm QCD}^2}$, where
$\Lambda_{\rm QCD}$ is the strong interaction scale. The precise value of 
$\Lambda_{\rm QCD}$ to be used here is difficult to estimate, but fortunately, 
these effects are suppressed by $1 - 4\sin^2\theta_W$. On the other hand, 
the relatively large $WW$-box contribution, $\Box_{WW} $, of about 26\% 
requires inclusion of two-loop mixed EW-QCD corrections, which in this case 
(and also for the $ZZ$-box, $\Box_{ZZ}$) are of short-distance type and given 
by~\cite{Erler:2003yk},
\begin{equation}
  \Box_{WW} = {\hat\alpha(M_W)\over 4\pi\sin^2\hat\theta_W(M_W)} 
              \left[ 2 + 5 \left( 1 - {\alpha_s(M_W)\over \pi} \right) \right].
\end{equation}

The two weak charges, $Q_W^p$ and $Q_W^e$, are complementary not only because 
of their very different experimental systematics, but also due to 
their different sensitivities to new physics~\cite{Erler:2003yk}. {\em E.g.}, 
supersymmetric loop contributions and many types of extra neutral $Z'$ bosons 
would affect them in a strongly correlated way. By contrast, SUSY models with 
so-called R-parity violation typically produce anti-correlated effects. 
And leptoquarks could strongly contribute to $Q_W^p$, but not to the purely 
leptonic electron weak charge. 

\subsection{APV}
Observations of atomic parity violation (APV) can be used to extract the weak
charges of heavy nuclei. These are defined analogous to $Q_W^e$ and $Q_W^p$, 
but come with entirely different experimental and theoretical issues. 
In particular, one needs a solid understanding of the structure of 
many-electron atoms~\cite{Flambaum:2006}. At present, only in 
$^{133}$Cs~\cite{Wood:1997zq,Guena:2005uj} and
$^{205}$Tl~\cite{Edwards:1995,Vetter:1995vf} are both the experimental and 
atomic theory errors at the $\%$-level, yielding 
$Q_W^{\rm Cs} = - 72.62 \pm 0.46$ and $Q_W^{\rm Tl} = - 116.4 \pm 3.64$, 
respectively. Future directions include measurements in Fr (using atom traps), 
and Ba$^+$ (which has a Cs-like atomic structure) may be studied in ion traps. 
An alternative could be the study of isotope ratios in which most of 
the atomic theory uncertainties cancel. Effects from the poorly known neutron 
distributions contribute an uncertainty at 
the $0.15\%$-level~\cite{Derevianko:2001uq}. This would be a problem for 
the isotope ratios unless our understanding of the neutron density can be 
improved. If this turned out to be impossible, one may conversely use APV 
to study nuclear structure. The weak charges of single isotopes (but not of 
isotope ratios) yield very different linear combinations of the coefficients 
$C_{1j}$ than $Q_W^p$, so that with Qweak it will be possible to constrain 
the individual $C_{1j}$ precisely. For a recent global fit to the $C_{ij}$, see
Ref.~\cite{Yao:2006px}.

\begin{figure}[t]
  \includegraphics[height=0.3\textheight]{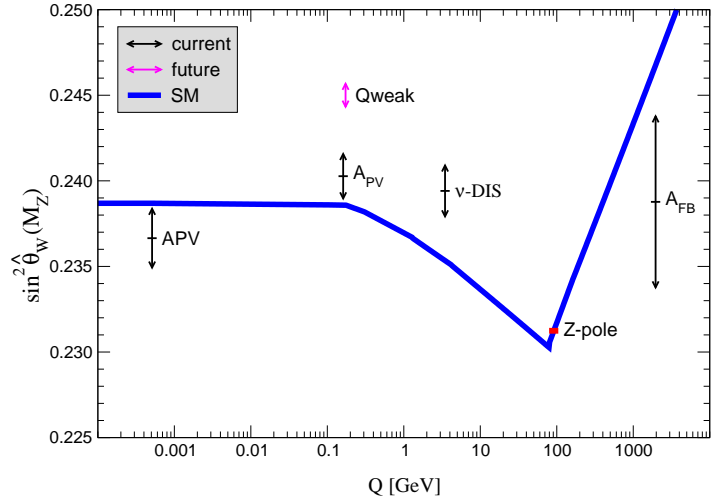}
  \caption{The weak mixing angle in the $\overline{\rm MS}$ renormalization 
scheme as a function of momentum transfer, $\sqrt{Q^2}$. The width of the line 
indicates the uncertainty in the SM prediction~\cite{Erler:2004in}. $A_{PV}$ is
the result from the M\o ller asymmetry~\cite{Anthony:2005pm}, $A_{FB}$ is 
the lepton forward-backward asymmetry from the Tevatron~\cite{Acosta:2004wq}.}
\label{fig:s2w}
\end{figure}

\subsection{NuTeV}
\label{sec:nutev}
The NuTeV experiment~\cite{Zeller:2001hh} in $\nu$-DIS finds for the on-shell 
definition of the weak mixing angle, $\sin^2 \theta_W = 0.2277\pm 0.0016$, 
which is 3.0~$\sigma$ higher than the SM prediction. The discrepancy is in 
the left-handed effective four-Fermi coupling, 
\begin{equation}
  g_L^2 = 0.3000 \pm 0.0014
  \sim \frac{1}{2} - \sin^2 \theta_W + \frac{5}{9} \sin^4\theta_W,
\end{equation}
which is 2.7~$\sigma$ low. Within the SM, one can identify five categories of 
effects that could cause or contribute to this effect: (i) an asymmetric 
strange quark sea, although this possibility is constrained by dimuon 
data~\cite{Goncharov:2001qe}; (ii) CSV PDFs at levels much stronger than 
generally expected~\cite{Martin:2003sk}; (iii) nuclear physics 
effects~\cite{Miller:2002xh,Kumano:2002ra,Brodsky:2004qa}; (iv) QED and 
electroweak radiative corrections~\cite{Arbuzov:2004zr,Diener:2003ss}; and 
(v) QCD corrections to the structure functions~\cite{Dobrescu:2003ta}. 
The NuTeV result and the older $\nu$-DIS data should therefore be considered as
preliminary until a re-analysis using PDFs including all experimental and 
theoretical information has been completed. It is well conceivable that various
effects add up to bring the NuTeV result in line with the SM prediction. It is 
likely that the overall uncertainties in $g_L^2$ (and $g_R^2$) will increase, 
but at the same time the older $\nu$-DIS results may become more precise when 
analyzed with better PDFs than were available at the time. The $\nu$-DIS 
results are compared with other determinations of $\sin^2 \theta_W$ in 
Fig.~\ref{fig:s2w}.

\subsection{$\mu$-decay}
The muon lifetime, $\tau_{\mu}$, yields a precise value for the Fermi constant
with negligible theoretical uncertainty, $G_F = 1.16637 \pm 0.00001$~GeV$^2$. 
There are two new efforts at PSI (FAST~\cite{Navarria:1998vz} and 
$\mu$LAN~\cite{Mulhauser:2006jd}) with the goal to improve $\tau_{\mu}$ by
a factor of 20 to $\sim 1$~ppm. The TWIST Collaboration at TRIUMF improved our 
knowledge of the model independent Michel parameters for $\mu$-decays. As shown
in Tab.~\ref{tab:Michel}, various parameters have already improved by about 
a factor of three, with further improvements expected.
\begin{table}
\begin{tabular}{cccrr}
\hline
\tablehead{1}{c}{b}{parameter} &
\tablehead{1}{c}{b}{comment}   &
\tablehead{1}{c}{b}{SM}        &
\tablehead{1}{c}{b}{pre-TWIST} &
\tablehead{1}{c}{b}{TWIST}     \\
\hline
$\rho$ & spectral shape & 3/4 & $0.7518\pm 0.0026$ & $0.7508\pm 0.0010$ \\
$\delta$ & asymmetry shape & 3/4 & $0.7486\pm 0.0038$ & $0.7496\pm 0.0013$ \\  
$P_\mu\zeta$ & asymmetry & 1 & $1.0027\pm 0.0085$ & $1.0003\pm 0.0038$ \\
$\eta$\tablenote{See~\cite{Fetscher:2006} for details.} &$m_e/m_\mu$-suppressed
& 0 & $-0.007\phantom{0}\pm 0.013\phantom{0}$ & $-0.0036\pm 0.0069$ \\
\hline
\end{tabular}
\caption{Results on the Michel parameters for muon decay.}
\label{tab:Michel}
\end{table}

\subsection{Muon Anomalous Magnetic Moment}
One of the most precisely measured observables is the anomalous magnetic moment
of the muon, $a_\mu$~\cite{Bunce:2006}. It is also easily affected by new 
physics contributions and therefore an important probe of physics beyond
the SM. However, the interpretation of $a_\mu$ is complicated by hadronic 
contributions. One can use $e^+ e^- \rightarrow$~hadrons cross section data 
to estimate the two-loop vacuum polarization (VP) effect. The most recent 
evaluation yields, 
$a_\mu^{(2,{\rm VP})} = (68.94\pm 0.46)\times 10^{-9}$~\cite{Hagiwara:2006jt},
implying a 3.4~$\sigma$ discrepancy between SM and 
experiment~\cite{Bennett:2004pv}. If one assumes isospin symmetry (which is not
exact and appropriate corrections~\cite{Cirigliano:2002pv} have to be applied) 
one can also make use of $\tau$ decay spectral functions~\cite{Schael:2005am} 
which yields instead~\cite{Davier:2003pw}, 
$a_\mu^{(2,{\rm VP})} = (71.10\pm 0.58)\times 10^{-9}$, and would remove
the discrepancy. It is not clear that the conflict between $e^+ e^-$ and $\tau$
data originates from larger-than-expected isospin violations:
Ref.~\cite{Maltman:2005yk} shows on the basis of a QCD sum rule that the $\tau$
decay data are consistent with values of 
$\alpha_s(M_Z) \buildrel > \over {_\sim} 0.120$ (in agreement with 
the result~(\ref{fit})), while the $e^+e^-$ data prefer lower (disfavored) 
values. Fortunately, as far as $a_\mu^{(2,{\rm VP})}$ is concerned, due to 
a suppression at large $Q^2$ (from where the conflict originates) this problem 
is less pronounced. An additional uncertainty is induced by hadronic three-loop
light-by-light-type graphs. A recent evaluation~\cite{Melnikov:2003xd} resulted
in $a_\mu^{\rm LBLS} = (1.36\pm 0.25) \times 10^{-9}$. This is higher than 
previous evaluations~\cite{Knecht:2001qf,Hayakawa:2001bb,Bijnens:2001cq}, but 
consistent with the simple quark level estimate of Ref.~\cite{Erler:2006vu}. 
The latter can also be used to bound $a_\mu^{\rm LBLS}$ from above, 
$a_\mu^{\rm LBLS} < 1.59 10^{-9}$ (95\% C.L.). If more experimental and 
theoretical work will be dedicated to these hadronic issues, a new and more 
precise experiment~\cite{Bunce:2006} of $a_\mu$ would very well be worth 
the effort.

\subsection{Electric Dipole Moments}
A very powerful probe of physics beyond the SM are searches for CP and time 
reversal symmetry (T) violating permanent electric dipole moments (EDMs) of 
electrons, muons, neutrons, and neutral atoms. EDM searches are of interest for
several reasons: (i) The SM (CKM) predictions for the magnitudes of EDMs fall 
well below the sensitivity of present and prospective measurements. 
Consequently, the observation of a non-zero EDM would signal the presence of 
physics beyond the SM or CP violation in the $SU(3)_C$ sector of the SM. 
The latter arises {\em via\/} a term in the Lagrangian~\cite{Wilczek:pj},
\begin{equation}
\label{eq:thetacp}
   {\cal L}_{\rm strong\ CP} = \theta_{\rm QCD} \frac{\alpha_s}{8\pi}
   G_{\mu\nu}{\tilde G^{\mu\nu}},
\end{equation}
where $G_{\mu\nu}$ (${\tilde G}_{\mu\nu}$) is the (dual) $SU(3)_C$ field 
strength tensor. (ii) The observed predominance of matter over anti-matter in 
the universe conflicts with expectations based on the SM alone. On the other 
hand, candidate extensions of the SM that could provide new CP violation of 
sufficient strength to account for the matter anti-matter asymmetry could also 
generate EDMs large enough to be seen. (iii) Recent experimental
developments~\cite{Flambaum:2006,Asahi:2006} have put the field on 
the verge of a revolution. The experimental sensitivities are poised to improve
by factors of 100 to 10,000 during the next decade. (iv) The various EDM 
searches provide complementary probes of new CP violation. {\em E.g.\/}, 
the observation of a non-zero neutron or atomic EDM in conjunction with a null 
result for the electron EDM at a comparable level of sensitivity would point 
toward the interaction of Eq.~(\ref{eq:thetacp}) as the likely source. 
In contrast, a non-zero lepton EDM would be a smoking gun for CP violation 
outside the SM, and a comparison with neutron and atomic studies would be 
essential for identifying the particular scenario responsible.

\subsection{Conclusions}
A network of high precision polarized electron scattering experiments will
study the TeV scale in a network of measurements, especially at JLab. Next 
generation $\mu$-decay experiments will improve the precision of the muon 
lifetime and are looking for deviations from the $V-A$ structure of the SM. 
The anomalous magnetic moment of the muon deviates at the 3~$\sigma$ level from
the SM prediction and is well worth further investments on both the theoretical
and experimental sides, considering that very many types of new physics 
scenarios can affect this observable. Searches for permanent EDMs are highly 
motivated, and is an area with spectacular experimental developments. 

I hope that this survey (although necessarily incomplete) serves 
to demonstrates that low energy measurements --- almost all of which using spin
degrees of freedom in an essential way --- will remain indispensable even in 
the LHC era.

\begin{theacknowledgments}
It is a great pleasure to thank the organizers for the kind invitation and for
an enjoyable and memorable conference. I am particularly grateful to Naohito 
Saito for his support and help. I also thank Paul Langacker and Michael 
Ramsey-Musolf for fruitful collaborations. This work was supported by CONACyT 
(M\'exico) contract 42026--F and by DGAPA--UNAM contract PAPIIT IN112902.
\end{theacknowledgments}

\end{document}